\documentclass{ws-procs9x6}
\def\be{\begin{equation}}
\def\ee{\end{equation}}
\def\bea{\begin{eqnarray}}
\def\eea{\end{eqnarray}}
\def\ket{\rangle}
\def\bra{\langle}
\begin{document}

\title{$3\pi$ photo-production with CLAS \footnote{\uppercase{T}his work is supported by the \uppercase{U}.\uppercase{S}. \uppercase{D}epartment of \uppercase{E}nergy and the \uppercase{U}.\uppercase{S}. \uppercase{N}ational \uppercase {S}cience \uppercase{F}oundation.}}

\author{M. Nozar for the CLAS Collaboration}

\address{ Thomas Jefferson National Accelerator Facility\\
12000 Jefferson Ave, MS 12H\\ 
Newport News, VA 23606\\ 
E-mail: nozarm@jlab.org}
\maketitle
\abstracts{
The $3\pi$ system produced in the reaction $\gamma p \rightarrow \pi^{+}\pi^{+}\pi^{-} n$ at $4.8\!-\!5.4$ GeV is investigated from the E01-017 (g6c) running of CLAS.  This energy range allows for the study of excited mesons in the $1\!-\!2$ GeV mass range, in their decay to $3\pi$, proceeding through $\rho \pi$ and $f_2 \pi$ emissions.  At these energies, there is significant overlap in phase space for events with $\rho$ and $f_2$ production, recoiling off an excited baryon, such as the $\Delta(1232)$, $N^*(1520)$ and $N^*(1680)$.  We show that after few kinematic selections, events of the latter type are suppressed in the final data set, allowing us to perform a PWA on the $3\pi$ system.}

\section{Introduction}\label{sec:Intro}
Due to the self interacting nature of gluons, QCD allows for hybrid states with a $(q\bar{q}g^{n})$ configuration, where the gluon excitation gives rise to a spectrum of additional states outside the constituent quark model.  One of the signature hadronic states is a meson with $J^{PC} = 1^{-+}$ quantum numbers which can not be attained by regular $(q\bar{q})$ mesons.  The inherent exotic quantum numbers prevent the mixing of this state with the conventional mesons with a $(q\bar{q})$ state configuration, thus simplifying the identification of such a state.  

There are several reasons behind selecting the charged $3\pi$ system for a Partial Wave Analysis (PWA).  The first reason is the simplicity of the final state, with only few decay channels open in the decay of the $3\pi$ system.  The limited acceptance of the CEBAF Large Acceptance Spectrometer (CLAS) for forward going particles associated with excited meson production poses a drawback for any meson spectroscopy experiment with CLAS in its current configuration; however, there is reasonable acceptance for detecting up to $3$ charged particles in CLAS.  Secondly, even though the dominant decay mode of this state is predicted to be through an $S$- and a $P$-wave meson emission, such as $b_1(1235)\pi$ and $f_1(1285)\pi$, the $\rho\pi$ decay channel is non negligible ~\cite{Close_Page}.  The evidence for the exotic $\pi(1600)$ state in the $\pi^{-} p \rightarrow \pi^{+}\pi^{-}\pi^{-} p$ reaction at $18$ GeV, by the Brookhaven E852 experiment~\cite{E852_3pi_1,E852_3pi_2}, provided yet another motivation to search for the state in the charged $3\pi$ system at JLab.  The production of the state was shown to be dominated by a natural parity (most likely a $\rho$) exchange.  In the framework of the Vector Dominance Model (VDM), with the photon beam turning into a vector meson, i.e. $\rho,\omega,\phi$, by reversing the role of the beam and the exchange particle ($Ex$), this state should also be produced with a photon beam in the pion exchange channel.  The are various discussions in the literature as to why photo-production may be a better production mechanism for exotic mesons~\cite{AA_AS_00,AA_PP_98,FC_PP_95}.  Photon beams, as probes for exotic meson production, have not been fully explored so far and the existing data on multi-particle final states are very sparse.   This experiment, with more statistics, should provide some guidance. 

There are three data sets of relevance to our analysis in photo-production: the SLAC 1-m. hydrogen bubble chamber experiment~\cite{Condo_91_93}, using a backscattered laser photon beam of $19.5$ GeV average energy from an incident electron beam of $30$ GeV; the CERN hydrogen experiment~\cite{Aston_81}, utilizing a tagged bremsstrahlung photon beam in the $25$-$70$ GeV energy range; and the SLAC 40-in. hydrogen bubble chamber experiment, using $4.3$ and $5.25$ GeV photon beams, produced by the $8.5$ and $10$ GeV positron annihilations in an $LH_2$ target~\cite{Eisenberg_Ballam_69}.  These experiments lacked the statistics required for a full PWA.

In Ref.~\refcite{Condo_91_93}, the analysis of the $\pi^{+}\pi^{+}\pi^{-}$ events in the reaction $\gamma p \rightarrow \pi^{+}\pi^{+}\pi^{-} n$ showed that the $3\pi$ spectrum in the low mass region is dominated by $a_2(1320)$ production with no clear evidence for $a_1(1260)$.  In the high mass region, based on the angular distribution analysis of the $3\pi$ events, the group claimed evidence for a narrow state at $1.775$ GeV with possible $J^{PC} = 1^{-+}, 2^{-+}$, or $3^{++}$ quantum number assignments.

From the analysis of the $4\pi$ events in $\gamma p \rightarrow \pi^{+}\pi^{-}\pi^{+}\pi^{-} p$, Ref.~\refcite{Aston_81}, reported two peaks, one at the mass of the $a_{2}(1320)$ and another at around $1.75$ GeV, in the $\rho \pi$ state recoiling off the remaining pion.  From the forward peaked nature of the $3\pi$ system, the production mechanism was attributed to the Deck-effect~\cite{Deck_64}.  No angular distribution analysis was performed on the data.

The authors of Ref.~\refcite{Eisenberg_Ballam_69} also reported two peaks in the $\rho \pi$ spectrum from the analysis of the $3\pi$ data in the $\gamma p \rightarrow \pi^{+}\pi^{+}\pi^{-} n$ reaction at the mass of the $a_2(1320)$ and one in the $\sim\! 1.7\!-\!1.85$ GeV region.  The production of the $a_{2}$ was shown to be consistent with a one-pion-exchange (OPE) mechanism.


In the case of charged $3\pi$ photo-production, the reaction is a charge-exchange process, and as such, neither Pomeron, nor $\omega$ exchanges are possible.  Considering $G$-parity conservation \footnote{since for a charged final state charge conjugation, $C$, is not a good quantum number}, pion as well as $a_1$ or $a_2$ are possible exchanges from the $\rho$ content of the photon beam, while for the $\omega$ part of the beam the most likely exchange is the $\rho$.  It is noteworthy that any contribution to Deck-effect enhancements~\cite{Deck_64} in either the $\rho^0 \pi^{+}$ or the $f_{2}\pi^{+}$ systems must come from the $\omega$ and $\phi$ components of the photon beam, due to $G$-parity conservation considerations.  

\section{Experimental Setup and Running Conditions}\label{sec:ExpSetup}
The data for this analysis were collected during Aug.--Sep. of $2001$.  The primary beam of $5.7$ GeV electrons at $100\%$ duty factor was provided by the Continuous Electron Beam Accelerator Facility (CEBAF).  The secondary beam of photons is produced in Hall B via bremsstrahlung radiation, using a radiator of $3\times10^{-4}$ radiation lenght.  The photon beam energy is determined by a tagging system which measures the momentum of the scattered electrons~\cite{TAGGER_NIM_00}.  The tagger is capable of identifying photons in the $20\%-95\%$ range of the incident electron beam energy.  An $18$ cm long cell filled with $LH_2$ was used as the proton target.  The Hall B houses the CLAS detector.  CLAS covers a large solid angle with polar angle detection in the range $8^{\circ} \leq\theta\leq 145^{\circ}$, and azimuthal angle coverage of $80\%$.  The detector, composed of six independent sectors, provides a toroidal magnetic field, where in normal settings, positively(negatively) charged particles bend outward(inward).  The three sets of drift chambers embedded in the space between the magnet coils in radial direction provide charged particle detection and track reconstruction.  A set of time of flight scintillators (TOF) are used for charged particle identification, and a set of electromagnetic calorimeters (EC) are used for neutral particle detection.  Further details of the CLAS detector design and performance are described elsewhere~\cite{CLAS_NIM_03}. 

To enhance the yield for the $\pi^{+}\pi^{+}\pi^{-}$ channel, the running conditions for $g6c$ were modified from the conventional photon beam runs at CLAS.  To increase the photon flux, the experiment ran with a higher electron beam intensity (with $\sim$$50\%$ of the data collected at $40$ nA and $\sim$$50\%$ at $50$ nA).  These electron beam intensities correspond to a photon beam flux of $\sim$$1.17\times10^{8}$ $\gamma/\rm sec$ and $\sim$$1.5\times10^{8}$ $\gamma/\rm sec$ in the entire tagging range, and $\sim$$8.8\times10^{6}$ $\gamma/\rm sec$ and $\sim$$1.1\times10^{7}$ $\gamma/ \rm sec$ in the top $15\%$ of the photon beam energy ($4.8$-$5.4$ GeV). To increase the acceptance for the negatively charged (inbending) particles, the target was pulled back by $100$ cm from the center of CLAS and the torus magnetic field was set to its half maximum value, corresponding to the torus current $I = 1938$ A.  

The level I trigger for the experiment was composed of a coincidence between a signal in the first $12$ tagger elements, a signal in $2$ of the $3$ Start Counter (ST) elements (an assembly of scintillators in three segments) and two charged particles in the TOF.  The level II trigger required two tracks in the drift chambers in any two sectors of CLAS.

\section{Event Selection}\label{sec:EvtSel}
In the $\pi^{+}\pi^{+}\pi^{-} (n)$ final state, the three pions were detected in CLAS and the neutron was reconstructed by missing mass.  Events which did not satisfy charge conservation in the reaction were rejected at the early stages of the analysis.  In addition, only events with two identified $\pi^{+}$, one $\pi^{-}$, and no more than two detected neutral particles were selected.  Vertex position cuts were applied to ensure the events originated within the target volume and a vertex timing constraint was imposed to reduce the accidental coincidences between the CLAS and the tagging system.  

The interaction of interest in our analysis is the $3\pi$ final state produced in $t-$channel exchange process, as shown in Fig.~\ref{fig:reaction_fg}.  With the maximum available photon beam energy of $5.4$ GeV, there is a non-negligible contribution from $t-$channel baryon resonance production from the two processes shown in Fig.~\ref{fig:reaction_bg}.  Of the two, the left process with the either the $\rho$ or the $f_2$ recoiling off a $\Delta(1232)/N^*$, is by far the largest.  In addition, observation of any features in the $n\pi\pi$ distribution required a selection around the $\Delta(1232)$ in the $n\pi$ distribution.  To enhance events from the process of interest, the photon beam energy was selected to be higher than $4.8$ GeV.  Furthermore, the peripherality condition was imposed by requiring that the four-momentum transfer squared from the photon to the $3\pi$ system, $-t'$, to be less than $0.4$ GeV$^2$.  In addition, since the two positively charged pions are the pions most likely to take part in the production of the baryon resonance, only forward-going $\pi^{+}$ were selected.  The cut was defined as $\theta_{lab}(\pi^{+})\leq 30^{\circ}$.  In the remainder of this report, we refer to the latter two cuts as the ``excited baryon rejection'' cuts. 
\begin{figure}[htb!]
\begin{minipage}[t]{0.28\linewidth}
\includegraphics[width=1.1in,totalheight=1.1in]{gamma_p_to_3pi_n.epsi}
\caption{Signal process: $t$-channel exchange $3\pi$ production.} \label{fig:reaction_fg}
\end{minipage}%
\hspace{0.15\textwidth}%
\begin{minipage}[t]{0.56\linewidth}
\includegraphics[width=1.1in,totalheight=1.1in]{gamma_p_to_2pi_npi.epsi}
\hspace{0.1\textwidth}%
\includegraphics[width=1.1in,totalheight=1.1in]{gamma_p_to_pi_npipi.epsi}
\caption{Background processes: $2\pi$ system recoiling off the $n\pi$ (left), Single $\pi$ recoiling off the $n\pi\pi$ (right).}
\label{fig:reaction_bg}
\end{minipage}
\end{figure}

\section{Data Distributions}\label{sec:DataDist}
\begin{figure}[hbt!]
\begin{minipage}{0.5\linewidth}
\includegraphics[height=2.2in,width=2.2in]{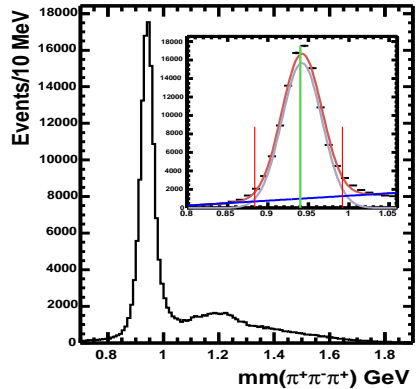}
\end{minipage}%
\hspace{0.05\textwidth}%
\begin{minipage}{0.45\linewidth}
\vskip-10mm
\caption{ The missing mass off the $\pi^{+} \pi^{+} \pi^{-}$ for low $-t'$ events ($-t' \leq 0.4$ GeV$^2$).  The first peak is at the mass of the neutron and the second peak is most likely associated with $n\gamma$ or $n\pi^{0}$ production.  The inset shows a Gaussian plus a first-order polynomial fit to the peak, giving a mass of $0.942$ GeV and $25$ MeV for $\sigma$.} \label{fig:mm}
\end{minipage}
\end{figure}

The missing mass off the $\pi^{+} \pi^{+} \pi^{-}$ for low $-t'$ events is showing in Fig.~\ref{fig:mm}.  The neutron peak sits on top of a linearly increasing background, with a signal to background ratio of approximately $9:1$.  The region between the lines, $(0.884 \leq mm \leq 0.992)$ GeV, indicates the neutron selection cut.  A Gaussian plus a $1^{\rm st}$ order polynomial fit to the distribution in this region, gives a mass of $0.942$ GeV and $25$ MeV for $\sigma$.

\begin{figure}[htb!]
\begin{minipage}[t]{0.5\linewidth}
\includegraphics[width=2.2in,totalheight=2.2in,clip=]{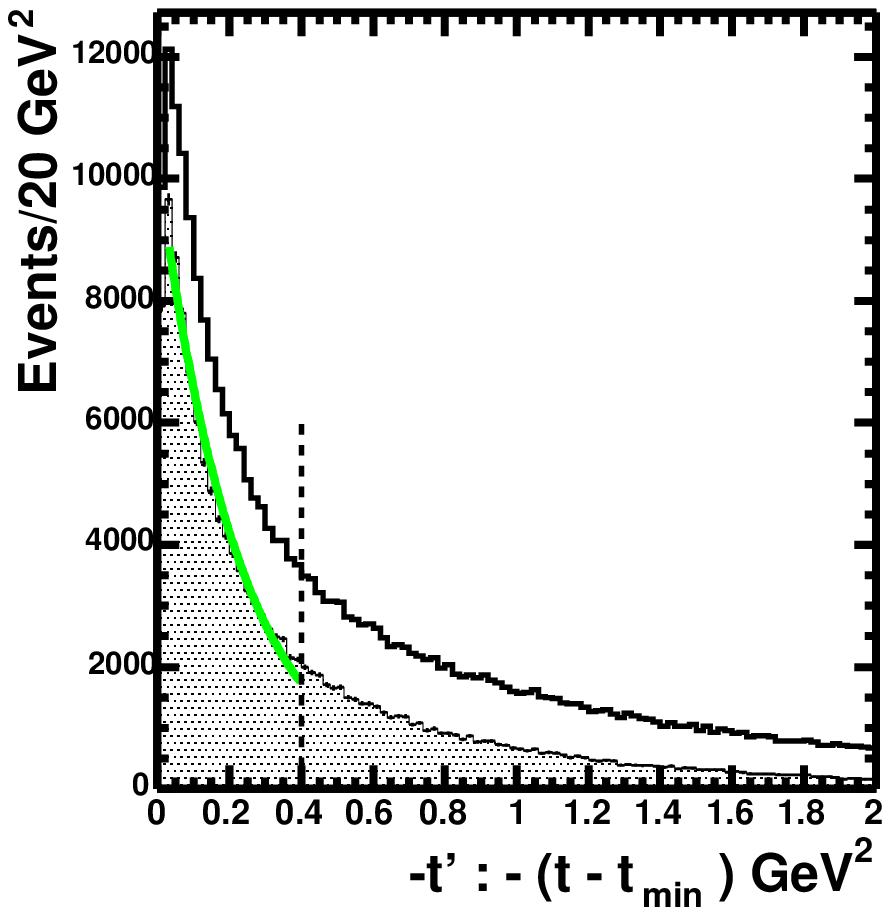}
\end{minipage}%
\begin{minipage}[t]{0.5\linewidth}
\includegraphics[width=2.2in,totalheight=2.2in,clip=]{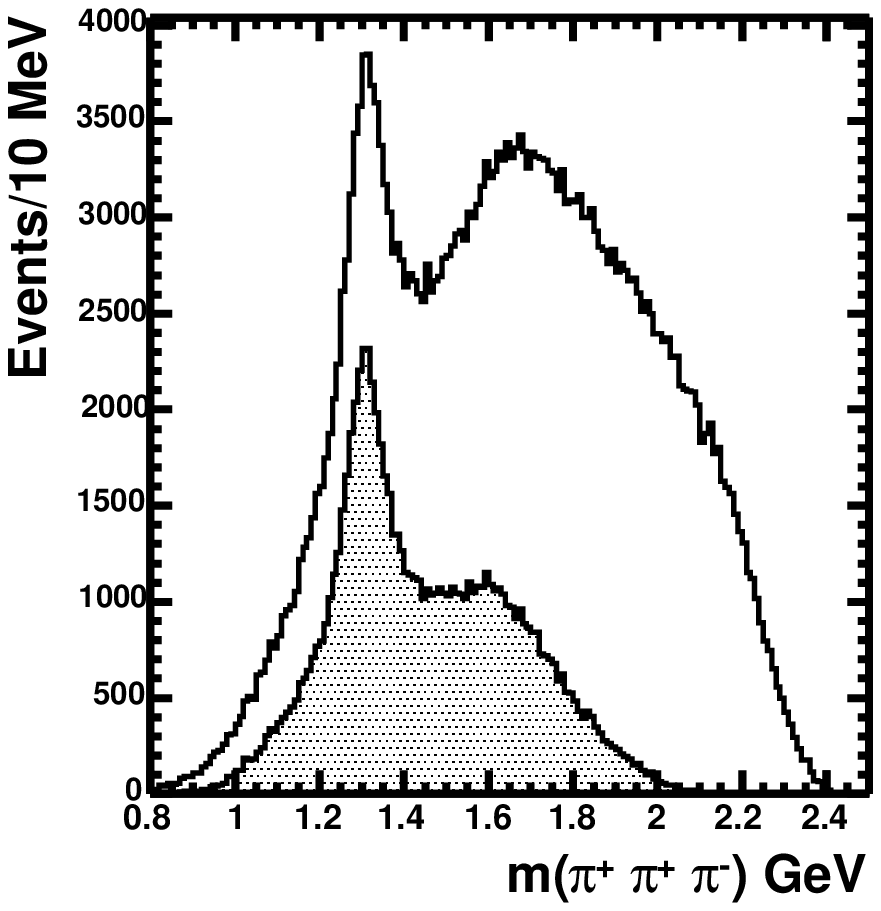}
\end{minipage}
\caption{Left: $t' = t-t_{min}$ from the beam to the $\pi^{+} \pi^{+} \pi^{-}$ system.  The shaded histogram shows the distribution after choosing forward-going positively charged pions.  The distribution is fit to an exponential of the form $a\,e^{-b|t'|}$, with $b=4.4$ GeV$^2$.  Right: $\pi^{+} \pi^{+} \pi^{-}$ invariant mass distribution.  The shaded histogram shows the distribution for events which passed the ``excited baryon rejection'' cuts, discussed in Sec.~\ref{sec:EvtSel}.} \label{fig:t_3pi}
\end{figure}

\vskip-5mm
The $-t'$ distribution, defined as $-t'=-(t-t_{min})$, with $-t$ the four-momentum transferred squared from the photon to the $3\pi$ system, is shown in the left plot of Fig.~\ref{fig:t_3pi}.  The shape of the distribution is consistent with the characteristics of peripheral production.  The distribution after the ``excited baryon rejection'' cuts, as defined in Sec.~\ref{sec:EvtSel}, is fit to an exponential function of the form $f(t') = a\,e^{-b|t'|}$, over the range $(0\leq -t' \leq 0.4)$ GeV$^2$.  The exponential constant, $b=4.4$ GeV$^{-2}$ is consistent with $\pi$ and $\rho$ exchange~\cite{MG_JM_MV_97}.  In the $3\pi$ invariant mass spectrum shown in the right plot of Fig.~\ref{fig:t_3pi} two enhancements are evident, one in the $1300$ MeV region, and another in the $1600\!-\!1700$ MeV mass range.
\vskip-3mm

\begin{figure}[htb!]
\includegraphics[width=4.65in,totalheight=3in]{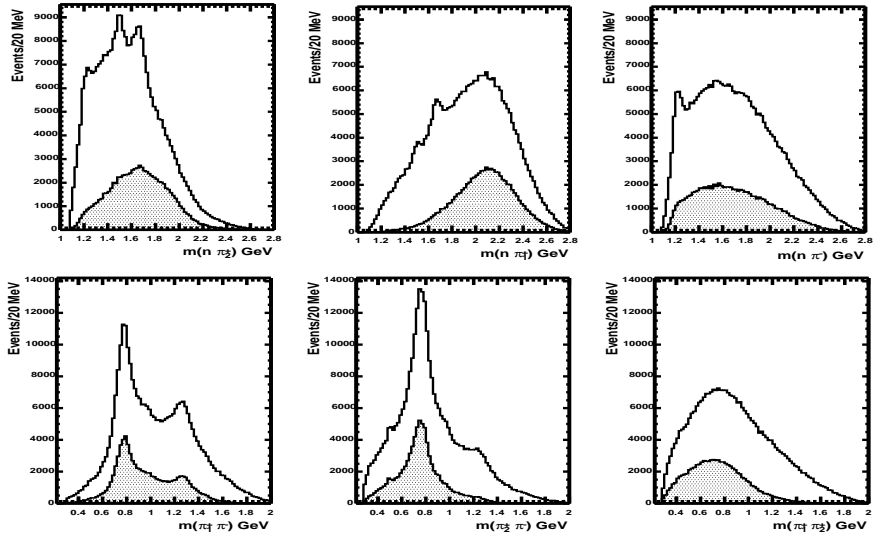}
\caption{The $n \pi$ (top row) and $\pi\pi$ (bottom row) invariant mass distributions.  The shaded histograms represent events which passed the ``excited baryon rejection'' cuts, defined in Sec.~\ref{sec:EvtSel}.}\label{fig:npi_pipi}
\end{figure}

Figure~\ref{fig:npi_pipi} shows all three possible combinations of the $n \pi$ and $\pi\pi$ invariant mass distributions.  In this analysis, the two positively charged pions were sorted based on momentum, with the $\pi^{+}_1$ being the more energetic of the two.  The two $n \pi^{+}$ combinations show peaks around the known baryon resonances, $\Delta(1232)$, $N^{\star}(1520)$, and $N^{\star}(1680)$, while the $n\pi^{-}$ shows a peak around the $\Delta(1232)$ only, as is expected due to isospin considerations.  It is clear from the shaded distributions that the baryon resonance peaks are significantly suppressed after the ``excited baryon rejection'' cuts.  The neutral $2\pi$ effective mass distributions show signals around the mass of the $\rho(770)$ and the $f_{2}(1270)$, as well as a shoulder at the mass of the $f_{0}(980)$.  The doubly-charged $2\pi$ combination does not show any distinct features, indicative of the lack of an isospin $I=2$ state.  
\vspace*{-2mm}

\section{Partial Wave Analysis}\label{sec:PWA}

The purpose of Partial Wave Analysis is to parameterize the observed intensity distribution in terms of a complete set of physically meaningful variables.  In the formalism used here, the set of variables are the physical intermediate states produced in the reaction.  This allows a direct extraction of the spin and parity of the states contributing to the total intensity and the determination of the resonance behavior and properties such as the mass, the width, and the decay properties of the produced states.  The details of the formalism and the code are discussed elsewhere\cite{SUC_PWA,JPC_DPW_PWA}. Here, we only mention the basic idea and the assumptions made in the procedure.

\subsection{PWA Formalism}\label{subsec:PWA_formalism}
In the analysis presented here, we have assumed that the production process (see Fig.~\ref{fig:reaction_fg}) is dominated by $t$-channel production of a $3\pi$ system with Reggeon exchanges between the photon beam and the proton target and that after the ``excited baryon rejection'' cuts, the background processes (see Fig.~\ref{fig:reaction_bg}) with either one or two pions recoiling off a $\Delta$/$N^{*}$ are significantly reduced.

\begin{figure}[ht!]
\begin{minipage}{0.45\linewidth}
\centerline{\epsfxsize=1.8in\epsfbox{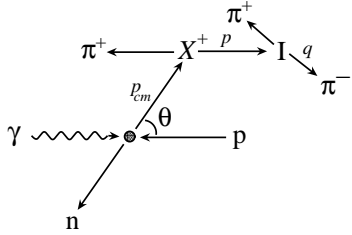}}   
\end{minipage}%
\begin{minipage}{0.55\linewidth}
\vskip-2mm
\caption{Photo-production of the $3\pi$ system, $X^+$, shown schematically in the C.M. frame,  with the sequential decay of $X^+$ to an isobar, $I$, and a $\pi$, in its rest frame, and the decay of $I$ into the remaining two pions.  $p_{cm}$ and $\theta$ represent the breakup momentum and polar angle of the $X$ in the C.M. frame.  $p$ represents the breakup momentum of $I$ in the $X$ rest frame, and $q$, the breakup momentum of one of the pions from the decay of $I$ in its rest frame.} \label{fig:gamma_p_to_X_n_cm}
\end{minipage}
\end{figure}

The reaction $\gamma p \rightarrow X^{+} n$, $X^{+} \rightarrow \pi^{+} \pi^{-} \pi^{+}$ is shown diagrammatically in Fig.~\ref{fig:gamma_p_to_X_n_cm}, with the production of $X^{+}$ in the Center of Mass (C.M.) frame, and its sequential decay to an isobar, $I$, and a $\pi$, followed by the decay of the isobar into the remaining $2\pi$.  The differential cross section for the reaction is given by:

\vskip-8mm
\be
\frac{d\sigma}{dcos(\theta)\; dM^{2}} = \int{|{\mathcal M}(\tau)|^2 d\rho(\tau)}
\ee

\noindent with $\theta$, the polar scattering angle of the $X^{+}$ in the C.M. frame (with $\hat z$ in the beam direction), $M$, mass of the $3\pi$ system, {$\mathcal M$}, the Lorentz-invariant amplitude, $d\rho(\tau) = p_{cm}\,d\tau$, the phase-space element with $p_{cm}$ the breakup momentum of the C.M. system and $\tau$ a set of five kinematic variables required to describe the $3\pi$ system.  For $\tau$ we have chosen $\Omega_{GJ}:(\theta_{GJ},\phi_{GJ})$, the Gottfried-Jackson angles describing the $X^{+} \rightarrow I \;\pi^{+}$ decay in the $X$ rest frame, $\Omega_h:(\theta_h,\phi_h)$, the helicity angles describing the $I \rightarrow \pi^{+} \; \pi^{-}$ decay in the Isobar rest frame, and $w$, the mass of the isobar.

\noindent For the PWA presented here, the data are binned in $t$ and the $3\pi$ mass, $M$.  Since $d\,t \propto p_{cm}\, d\, cos(\theta)$, the differential cross-section in a given $t$ and $M$ bin can be written as:
\be
\frac{d\sigma}{dt dM^{2}} = \left(\frac{1}{p_{cm}} \right)\int | {\mathcal M}(\tau)|^2 d\rho(\tau)
\ee
 
\noindent The intensity distribution in terms of {$\mathcal M$} is given by:
\be
I(\tau) \propto M |\; {\mathcal M}(\tau)|^2
\ee

\noindent Once we have assumed an interaction mechanism, i.e. in our case, $t$-channel $3\pi$ production through reggeon exchange between the photon beam and the proton target, we can write {$\mathcal M$} in terms of the transition operator, $\hat T$:
\be
{\mathcal M} = \bra 3\pi |\, \hat{T} \,| \gamma \, Ex \ket 
\ee

\noindent The above transition matrix element can be re-written in terms of production of intermediate mesonic states, $X$, each with a unique set of quantum numbers, $I^G J^{PC}\, m$, where $I$ denotes the isospin, $G$ the $G$-parity, $J$ the total spin, $P$ the parity, $C$ the charge-conjugation, and $m$ the $z$-projection of the spin (chosen along the beam direction).  Quantum-mechanically, this corresponds to expanding $\mathcal M$ in terms of a complete set of states, $| X \ket$, with $\sum_{X} |X \ket\bra X |= 1$:

\be
{\mathcal M} = \bra 3\pi |\, \hat{T}\,  \sum_{X} |X \ket\bra X| \gamma \; Ex \ket 
\ee

\noindent Assuming that $\hat{T}$ is separable into two parts, $\hat{T_p}$ and $\hat{T_d}$, where $\hat{T_p}$ and $\hat{T_d}$ are the production and decay operators for the intermediate states, $X$, then {$\mathcal M$} can be re-written:

\be
{\mathcal M} = \bra  3\pi | \, \hat{T_d}\, \sum_{X} |X \ket\bra X| \,\hat{T}_{p}\,| \gamma \; Ex\ket  = \sum_{X}\bra 3\pi|\, \hat{T_d}\,|X \ket\bra X|\, \hat{T}_{p}\,| \gamma\; Ex \ket   
\ee

\noindent The PWA formalism adopted here is based on the isobar model~\cite{Isobar_Herndon}, where the decay process of $X$ to $3\pi$ is described through a series of sequential two-body decays as shown in Fig.~\ref{fig:gamma_p_to_X_n_pwa}:
\begin{figure}[ht!]
\begin{minipage}{0.45\linewidth}
\centerline{\epsfxsize=1.8in\epsfbox{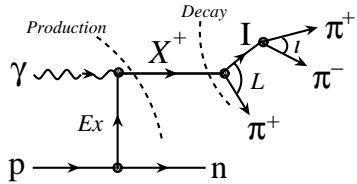}}
\end{minipage}%
\begin{minipage}{0.55\linewidth}
\vskip-2mm
\caption{$t$-channel exchange production of the possible $3\pi$ states, $X^{+}$, with their subsequent sequential decay, $X^{+}\rightarrow I \, \pi^{+}$ followed by the isobar decay, $I \rightarrow \pi^{+} \pi^{-}$.}\label{fig:gamma_p_to_X_n_pwa}
\end{minipage}
\end{figure}


\noindent The inclusion of all possible isobars in the decays in:  
\be
X^{+} \rightarrow I \; \pi^{+}_{1\,(2)} \;\; \; , \;\;\; I \rightarrow  \pi^{+}_{2\,(1)} \;\pi^{-}
\ee
\noindent is handled through the addition of a set of intermediate isobar states, $|I \ket$, in the calculation of the decay amplitudes:

\be
{\mathcal M} =  \sum_{X,I} \underbrace{\bra (\pi\pi)\;\pi | \,\hat{T}^{I\rightarrow \pi\;\pi}_d  \,|I\,\pi \ket\bra I\,\pi|  \,\hat{T}^{X\rightarrow I\; \pi}_d  \,|X \ket }_{Decay: A(\tau)}\underbrace{\bra X| \, \hat{T}_p \, |\gamma \; Ex \ket}_{Production: V}  
\ee
\noindent Then $I(\tau)$ can be written as:
\be
I(\tau) \propto  \sum_{k} {| \sum_{\alpha}{V_{k\alpha}A_\alpha(\tau)} |\, ^2} 
\ee

\noindent Here $k$ is the number of possible external spin configurations which in the case of an unpolarized photon beam and target, can be up to $8$; $\alpha:\{J,P,C,m,L,l,(w,\Gamma)\}$ is the set of quantum numbers describing a given state and its decay (as mentioned earlier, $J$, $P$, and $C$ are the $J^{PC}$ quantum numbers of $X$, and $m$, the $z$-projection of its spin, $J$); $L$ and $l$ are the orbital angular momenta between $I$ and $\pi$ in $X \rightarrow I \; \pi$ decay, and between the two pions in $I \rightarrow \pi \; \pi$ decay; $w$ and $\Gamma$ are the mass and the width of the isobars used in their Breit-Wigner description.  The angular dependencies of the decays are handled using the Wigner D-functions and the appropriate Blatt-Weisskopf angular momentum barrier factors are used for a given $L$ and $l$ involved in a decay chain.  The decay amplitudes are constructed as eigenstates of reflectivity to take advantage of the parity conservation in the production process~\cite{SUC_Truman}.  This choice reduces the possible number of external spin configurations by a factor of $2$ and reduces the spin-density matrix to a block-diagonalized form, where there is interference only between amplitudes of the same reflectivity, $\epsilon$.  
\be
I(\tau) = \sum_{k} \left\{ {\left| \sum_{\epsilon=+1} {^\epsilon V_{k\alpha}} {^\epsilon \!A_{\alpha}}(\tau) \right| }^2 + \, {\left|\sum_{\epsilon=-1} {^\epsilon V_{k\alpha}} {^\epsilon\!A_{\alpha}}(\tau) \right|}^2 \right\}
\ee

To see how well the description given above fits the data, event-based maximum likelihood fits are used in $40$ MeV mass bins of $3\pi$ in the $t$ range $0 \leq t \leq 0.4$ GeV$^2$. \\

The extended likelihood function is written as the product of the probabilities for finding $n$ events in a $3\pi$ mass bin:
\be
{\mathcal L} \propto \left[ \frac{{\bar n}^n}{n!} e^{-\bar{n}} \right]\prod_{i}^{n} \left[\frac{I(\tau_i)}{\int I(\tau)\, \eta(\tau)\, p q\, d\tau}\right] 
\label{eqn:LF}
\ee

The normalization term in the denominator of Eq.~\ref{eqn:LF} is determined through the calculation of normalization integrals over a mass bin.  The finite experimental acceptance of the detector, $\eta(\tau)$, is determined by the Monte Carlo method, where a set of $3\pi$ events in the $1-2$ GeV mass range with the same $t$ distribution as the final data set were generated according to phase space in $20$ MeV wide bins.  The number of generated events in each bin was chosen such that the number of final accepted Monte Carlo events were $15$ times greater than for the data in a given bin.  The average acceptance as a function of $3\pi$ mass is a smoothly varying distribution and is on the order of $4\%$.  The Monte Marlo events were subjected to the same analysis cuts as the data.  

For practical reasons $-{\rm ln}(\mathcal L )$ is minimized\footnote{rather than maximizing the $\mathcal L$.} by varying the production amplitudes as parameters in the fit, Eq.~\ref{eqn:LF} is rewritten as:
\bea
{\rm ln}(\mathcal L) &\propto& \sum_{i}^{n} {\rm ln}\, I(\tau_i) - \int  I(\tau) \eta(\tau)\, p\,q\, d\tau \\
             &\propto & \sum_i^n \ln \left[\, \sum_{k\epsilon \alpha\alpha'} {^\epsilon V}_{\alpha k} {^\epsilon V_{\alpha' k}^*} {^\epsilon\!A_{\alpha}}(\tau_i) {^\epsilon\!A_{\alpha'}^*}(\tau_i)\right] -\eta_x \left[\, \sum_{k\epsilon\alpha\alpha'} {^\epsilon V_{\alpha k}} {^\epsilon V_{\alpha' k}^*}  {^\epsilon\!\Psi_{\alpha\alpha'}^a} \right] \nonumber \,,
\eea
\noindent where $\eta_x = \frac{M_a}{M_r}$ is the M.C. acceptance with $M_a$ and $M_r$ the number of accepted and raw M.C. events in a given $3\pi$ mass bin and $^\epsilon\!\Psi_{\alpha\alpha'}^a$ represents the normalization integral, calculated for the accepted M.C. sample, defined as:
\be
{^\epsilon\!\Psi_{\alpha\alpha'}^a} = \frac{1}{M_a} \sum_{i}^{M_a} {^\epsilon\!A_{\alpha}}(\tau_i) {^\epsilon\!A_{\alpha'}^*}(\tau_i)
\ee

The number of events as predicted by the fit is given by:
\be
N = \sum_{k\epsilon \alpha\alpha'} {^\epsilon V}_{\alpha k} {^\epsilon V_{\alpha' k}^*} {^\epsilon\!\Psi_{\alpha\alpha'}^r}  \nonumber \,,
\ee  

\noindent with ${^\epsilon\!\Psi_{\alpha\alpha'}^r}$, the normalization integral calculated for the raw M.C. sample in a given $3\pi$ mass bin.

\subsection{Preliminary PWA Results}\label{subsection:PWA_res}
The PWA results shown here are still very preliminary.  In the choice of waves included in the fits we have taken a ``minimalistic'' approach, where only a minimal set of states are included in the fits.  The list of $35+1$ waves used in the PWA fit for which we present results here, is shown in Table~\ref{ListOfWaves}.  Since the partial waves are represented by complex amplitudes, $35$ waves corresponds to $70$ parameters plus one parameter for the non-interfering Background term.
\vskip-10mm
\begin{center}
\begin{table} [htb!]
\tbl{Set of partial waves used in the PWA fit.\vspace*{1pt}}
{\footnotesize
\begin{tabular}{cllcc}
\hline\\
$J^{PC}$ & $m^{\epsilon}$ & L & Isobar & $\#$ Waves\\ \hline

$0^{-+}$ & $0^{+}$                         & $0$      &  $\sigma$       & $1$   \\
$0^{-+}$ & $0^{+}$                         & $0$      &  $f_{0}(980)$   & $1$   \\  
$0^{-+}$ & $0^{+}$                         & $1$      &  $\rho(770)$    & $1$   \\[1ex] 

$1^{++}$ & $0^{+}$, $1^{\pm}$              & $0$,$2$  & $\rho(770)$     & $6$   \\
$1^{++}$ & $0^{+}$, $1^{\pm}$              & $1$      & $\sigma$        & $3$   \\ [1ex]

$1^{-+}$ & $0^{-}$, $1^{\pm}$              & $1$      & $\rho(770)$     & $3$   \\ [1ex]

$2^{++}$ & $0^{-}$, $1^{\pm}$, $2^{\pm}$    & $2$      & $\rho(770)$     & $5$   \\ [1ex]

$2^{-+}$ & $0^{+}$, $1^{\pm}$              & $1$,$3$  & $\rho(770)$     & $6$  \\ 
$2^{-+}$ & $0^{+}$, $1^{\pm}$              & $2$      & $\sigma$        & $3$  \\ 
$2^{-+}$ & $0^{+}$, $1^{\pm}$              & $0$,$2$  & $f_{2}(1270)$   & $6$  \\ [1ex]
Background & & & & $1$  \\ \hline
\end{tabular}
\label{ListOfWaves}}
\vspace*{-13pt}
\end{table}
\end{center}

Fig.~\ref{fig:PWA_results} shows the intensity distributions for various wave sets, as extracted from PWA.  Each point on the plots is a result of an independent fit.  The fit is a rank $1$ fit.  The strongest signal observed is in the $2^{++}$ intensity, at the mass of the $a_2(1320)$.  The width of the signal ($\sim 120$ MeV) is in agreement with the PDG value.  The next strongest signals are observed with approximately the same strength in the $1^{++}$ and $2^{-+}$ intensities.  The $1^{++}$ intensity shows an enhancement at the mass of the $a_1(1260)$, and the $2^{-+}$ intensity shows strength at the mass of $\pi_2(1670)$.  The $0^{-+}$ intensity shows some enhancement around the region where the $\pi(1800)$ is expected.  The exotic $1^{-+}$ wave intensity shows a peak at approximately $1700$ MeV with a narrow width of $\sim$$160$ MeV.  
\begin{center}
\vskip-7mm
\begin{figure}[htt!]
\begin{minipage}{0.46\linewidth}
\resizebox{12pc}{!}{\includegraphics{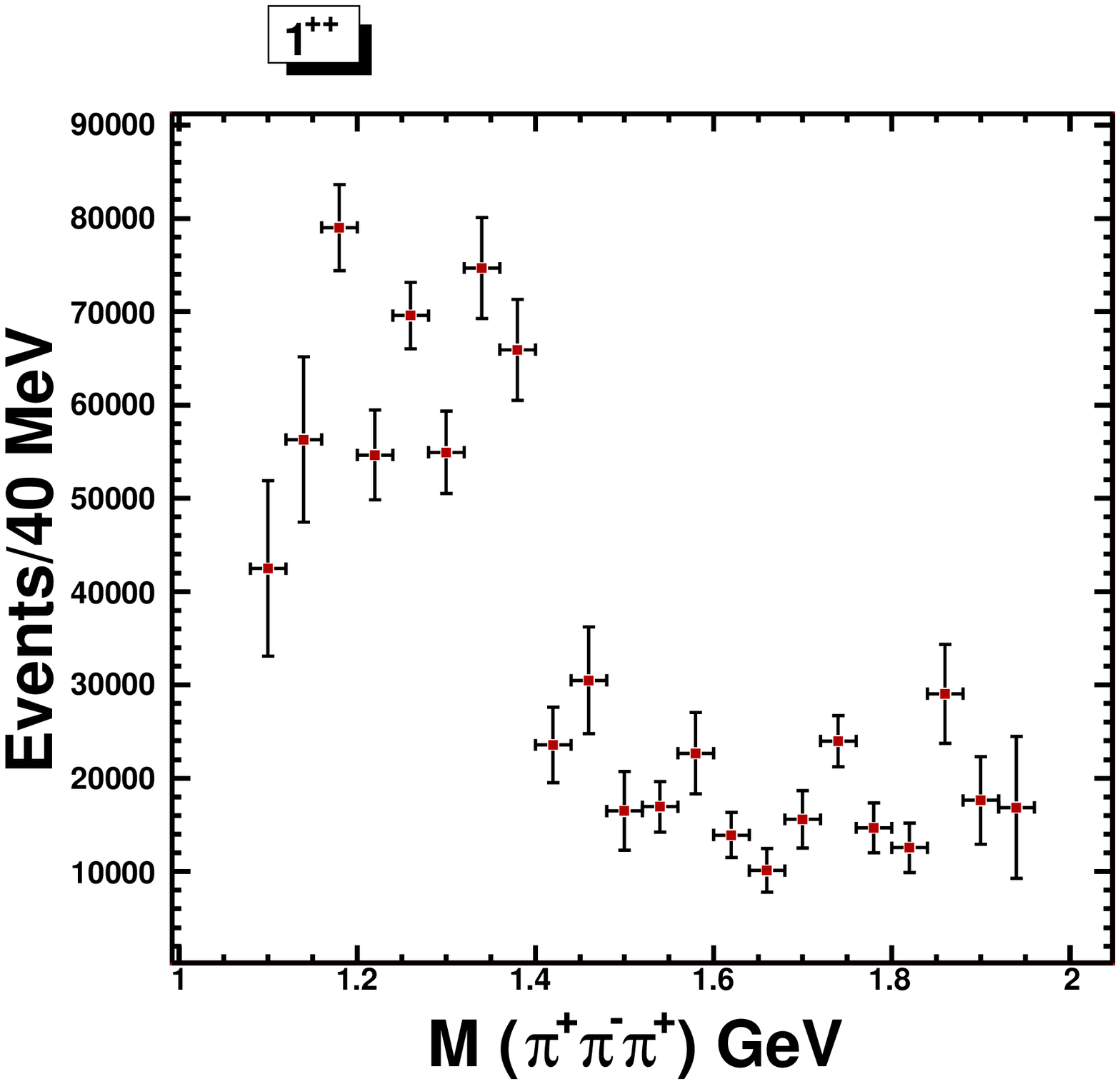}}
\end{minipage}%
\hspace{0.06\textwidth}%
\begin{minipage}{0.46\linewidth}
\resizebox{12pc}{!}{\includegraphics{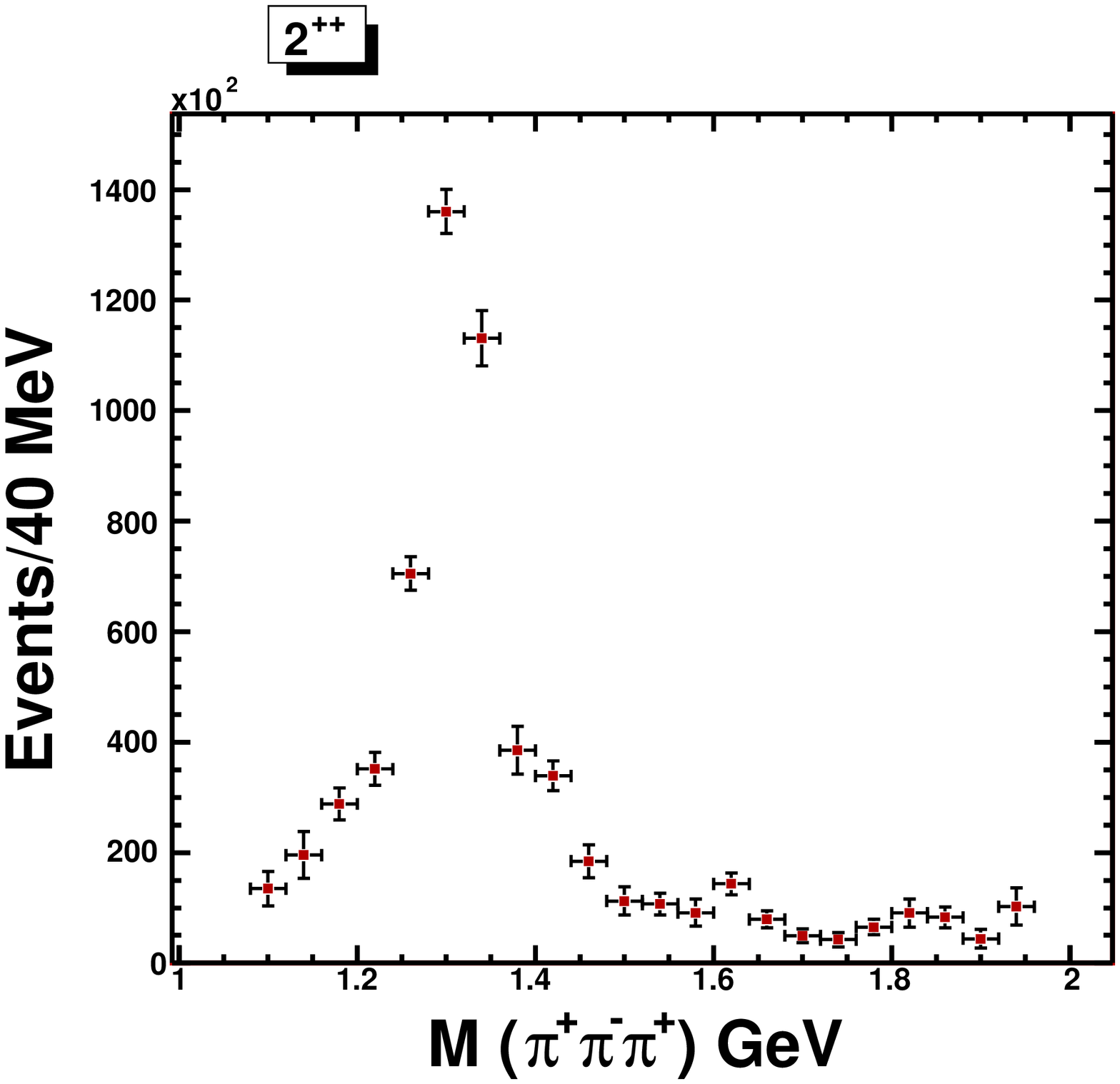}}
\end{minipage}\\
\begin{minipage}{0.46\linewidth}
\resizebox{12pc}{!}{\includegraphics{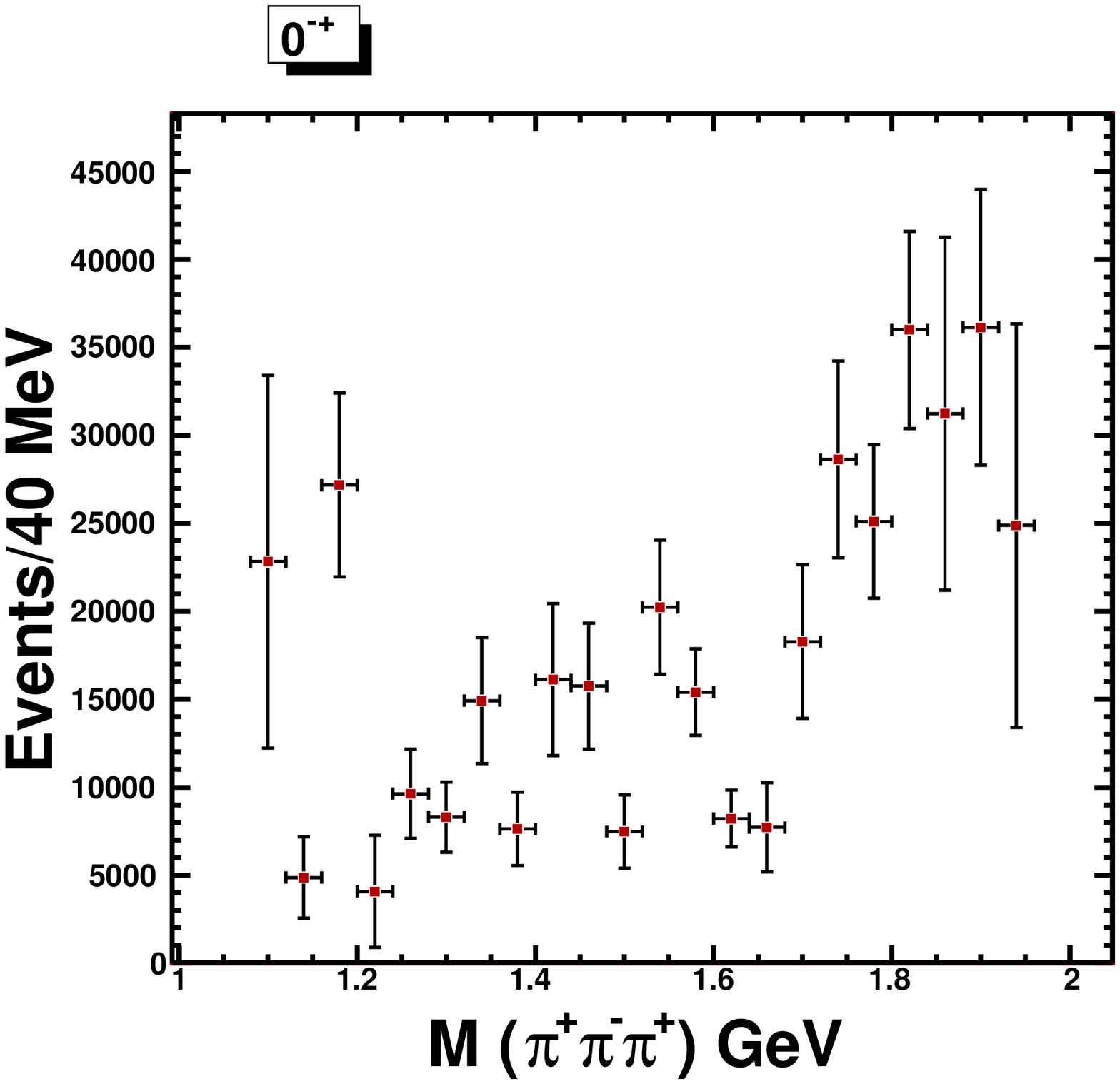}}
\end{minipage}%
\hspace{0.06\textwidth}%
\begin{minipage}{0.46\linewidth}
\resizebox{12pc}{!}{\includegraphics{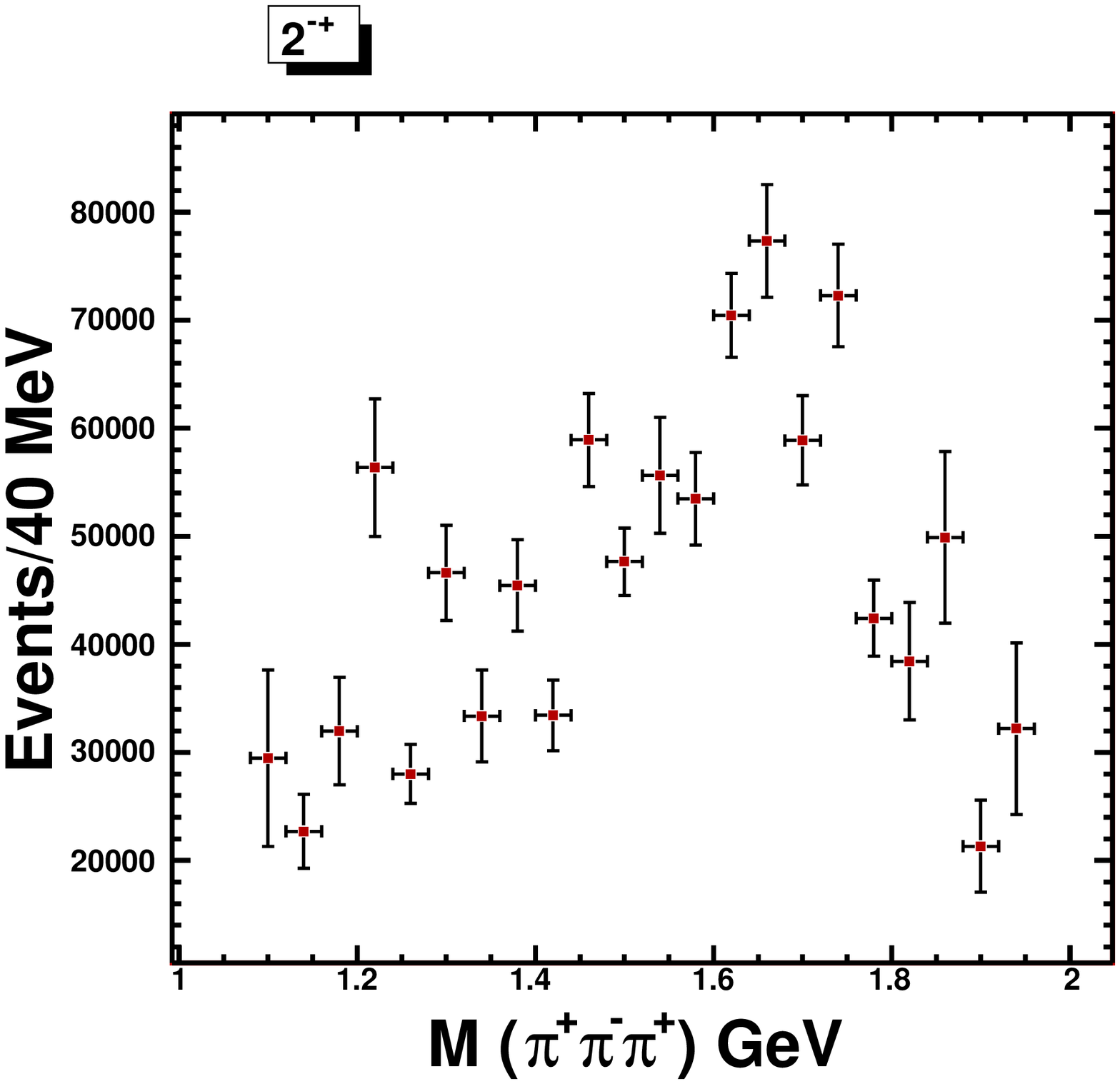}}
\end{minipage}\\
\hspace{0.06\textwidth}%
\begin{minipage}{0.46\linewidth}
\resizebox{12pc}{!}{\includegraphics{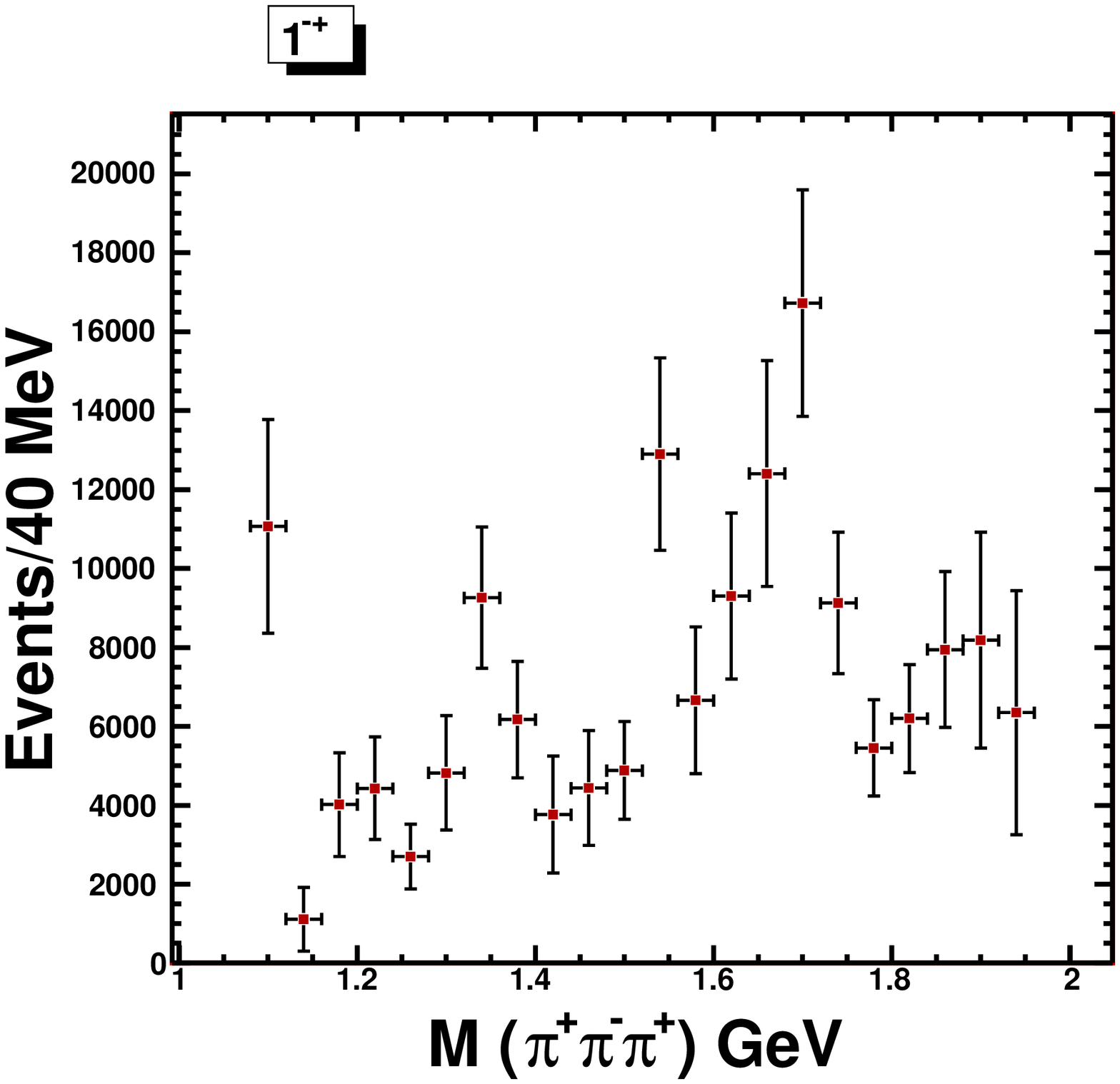}}
\end{minipage}%
\hspace{0.06\textwidth}%
\begin{minipage}{0.46\linewidth}
\vskip-10mm
\caption{PWA results: Combined intensities of all waves included in the fit.  Top: $1^{++}$ (left), $2^{++}$ (right); Middle: $0^{-+}$ (left), $2^{-+}$ (right); Bottom: $1^{-+}$.}\label{fig:PWA_results}
\end{minipage}
\end{figure}
\end{center}

\begin{figure}[htb]
\begin{minipage}{\linewidth}
\includegraphics[width=4.55in,height=2in,clip=]{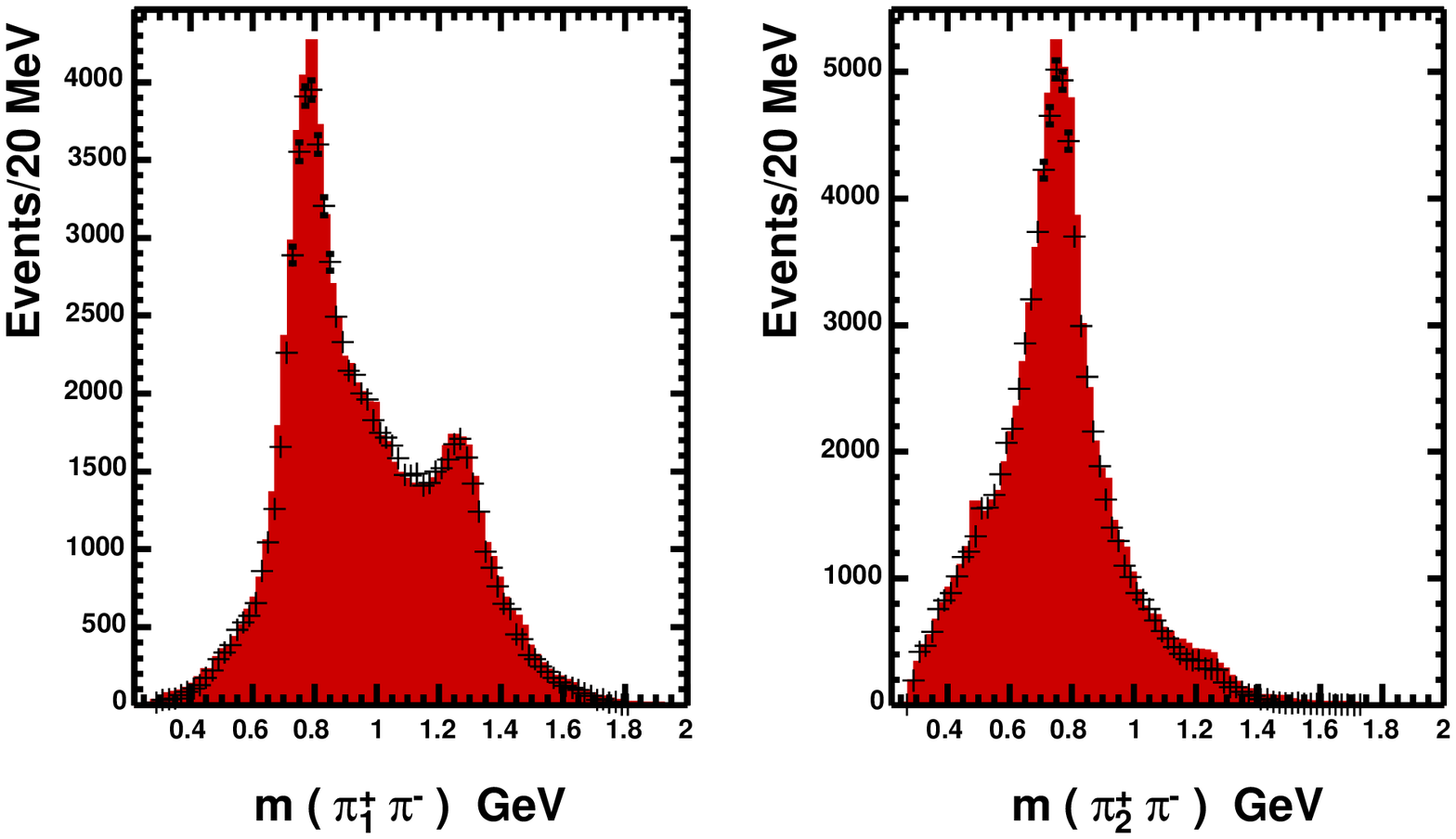}
\end{minipage}\\
\begin{minipage}{\linewidth}
\includegraphics[width=4.65in,height=2in,clip=]{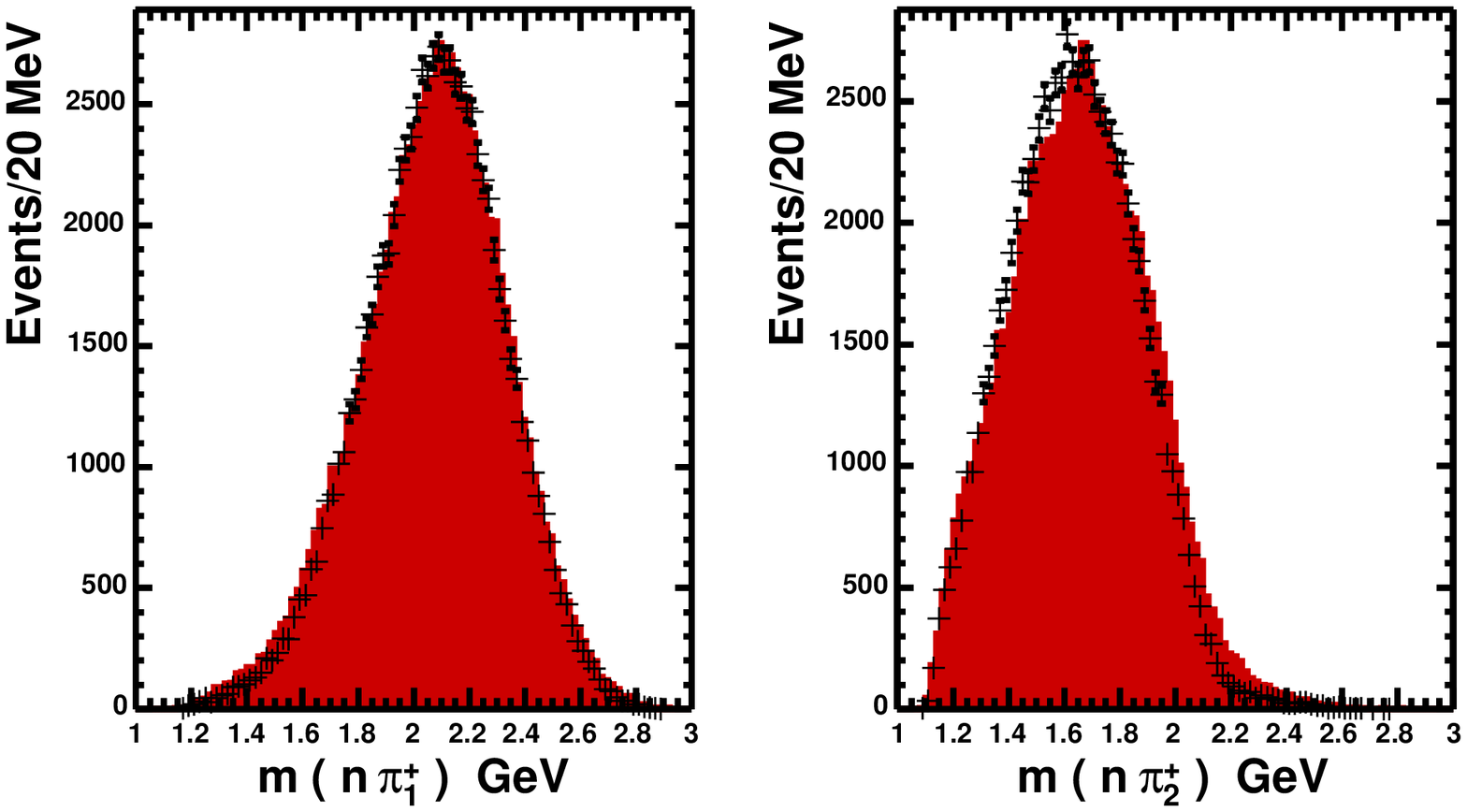}
\end{minipage}
\caption{Quality of the PWA fit results.  Comparison of various distributions between the ``experimental data'' (red shaded histograms) and the ``predicted data'' (black points) as determined by PWA.  Top: $\pi\pi$ invariant mass distributions.  Bottom: $n\pi$ invariant mass distributions.}\label{fig:predict}
\end{figure}

The quality of the PWA fit results are determined by comparing various data distributions with the data set as predicted by the PWA fit results.  The predicted data set are obtained by weighting the accepted phase space Monte Carlo events by the results of the PWA.  In Fig.~\ref{fig:predict} we show the $2\pi$ and $n\pi$ distributions for the data (red shaded histograms) and the PWA predicted data set (black points).  As can be seen, there is a good qualitative agreement between the fit results and the data.  There is some disagreement between the two data sets in the $n\pi^{+}_{2}$ distribution.  Since $\pi^{+}_{2}$ is the lower momentum of the two positively charged pions, it is more likely to be associated with the baryon resonance production in the lower vertex.  The disagreement, is therefore, indicative of the level of this background (i.e. remaining $\Delta/N^*$ after ``excited baryon rejection'' cuts.

\section*{Conclusions}
We have performed a PWA on a sample of $\sim$$84,000$ $\pi^{+}\pi^{+}\pi^{-}$ events from the $g6c$ experiment at CLAS. The PWA fit results are very encouraging but not finalized.  From the results shown in Fig.~\ref{fig:PWA_results} we can see fluctuations from bin to bin in the fit results.  This could be an indication that for some bins, we do not have the fit with the best likelihood value, but results from one of the local minima.  To remedy the situation, we will perform many fits in a given bin for the same set of input waves, but different random starting values for the parameters and look for the best solution.  We have also recently performed ``tracking'' fits where the results for the parameters in a given bin are used as the starting values for the fit in the adjacent bin.  We see improvement in the continuity of the fit results from bin to bin after tracking.  The number of waves used in the lower $3\pi$ mass region can be reduced further by using a different set of input waves for the low $(1.0-1.4)$ GeV and high $(1.4-2.0)$ GeV $3\pi$ mass region.  For instance, all $f_2(1270) \pi$ waves can be eliminated for fits below the $1.4$ GeV mass range due to threshold considerations. We will also try different modeling of the background term which at the present is included as a non-interfering isotropic wave.  A Mass Dependent (M.D.) fit to the results of the final PWA fits will allow us to extract resonance parameters of the observed states.
\section*{Acknowledgments}
The invaluable efforts of the Jefferson Lab Accelerator staff, the Physics Division staff, and the g6c group are greatly appreciated. TJNAF is operated by the Southeastern Universities Research Association (SURA) for the United States Department of Energy under contract $\#$ DE-AC0584ER40150.  This work was supported in part by the National Science Foundation (NSF) grant $\#$ 9733999. 



\begin{thebibliography}{0}

\bibitem{E852_3pi_1} G. S. Adams et al., {\it Phys. Rev. Lett.} {\bf 81}, 5760 (1998).
\bibitem{E852_3pi_2} S. U. Chung et al., {\it Phys. Rev.} {\bf D65}, 072001 (2002).
\bibitem{AA_AS_00} A. V. Afanasev and A. P. Szczepaniak, {\it Phys. Rev.} {\bf D61}, 114000 (2000).
\bibitem{AA_PP_98} A. V. Afanasev and P. R. Page, {\it Phys. Rev.} {\bf D57}, 67711 (1998).
\bibitem{FC_PP_95} F. E. Close and P. R. Page, {\it Phys. Rev.} {\bf D52}, 1706 (1995).
\bibitem{Close_Page} F. E. Close and P. R. Page, {\it Nucl. Phys.} {\bf B443}, 233 (1995).
\bibitem{Condo_91_93} Condo et al., {\it Phys. Rev.} {\bf D48}, 3045 (1993), {\it Phys. Rev.} {\bf D43}, 2787 (1991).
\bibitem{Aston_81} Aston et al., {\it Nucl. Phys.} {\bf B189}, 15 (1981).
\bibitem{Eisenberg_Ballam_69} Eisenberg et al., {\it Phys. Rev. Lett.} {\bf 23}, 1322 (1969).
\bibitem{Deck_64} R. T. Deck, {\it Phys. Rev. Lett.} {\bf 33}, 169 (1964).
\bibitem{TAGGER_NIM_00} D. I. Sober et al., {\it Nucl. Inst. and Meth. in Phys. Res.} {\bf A440}, 263 (2000).
\bibitem{CLAS_NIM_03} B. A. Mecking et al., {\it Nucl. Inst. and Meth. in Phys. Res.} {\bf A503}, 513 (2003).
\bibitem{MG_JM_MV_97} M. Guidal, J. M. Laget, M. Vanderhaeghen, {\it Nucl. Phys.} {\bf A627}, 645 (1997).
\bibitem{SUC_PWA}  S. U. Chung,
``Spin Formalism'', CERN Yellow Report CERN 71-8 (1971) , ``Formalism for Partial Wave Analysis, Version II'', BNL preprint QGS-93-05.

\bibitem{JPC_DPW_PWA} J. P. Cummings and D. P. Weygand, submitted to {\it Nucl. Inst. and Meth. in Phys. Res.} (2003).
\bibitem{Isobar_Herndon} D. J. Herndon and P. S\"{o}ding and R. J. Cashmore, {\it Phys. Rev.} {\bf D11}, 3165 (1975).
\bibitem{SUC_Truman} S. U. Chung, T.L. Trueman, {\it Phys. Rev.} {\bf D11}, 633 (1975).

\end{thebibliography}
\end{document}